\documentclass[conference]{IEEEtran}
\IEEEoverridecommandlockouts
\usepackage{amsmath,amssymb,amsfonts}
\usepackage{algorithmic}
\usepackage{graphicx}
\usepackage{textcomp}
\usepackage{xcolor}
\usepackage[numbers,sort&compress]{natbib}
\usepackage{glossaries}
\newglossaryentry{cpps}{
	name = {CPPS},
	description = {cyber-physical production systems},
	text = {CPPS},
	first = {cyber-physical production systems (CPPS)},
	plural = {CPPSs},
	firstplural = {cyber-physical production systems (CPPSs)}
}

\newglossaryentry{cps}{
	name = {CPS},
	description = {cyber-physical systems},
	text = {CPS},
	first = {cyber-physical systems (CPS)},
	plural = {CPSs},
	firstplural = {cyber-physical systems (CPSs)}
}

\newglossaryentry{re}{
	name = {RE},
	description = {reverse engineering},
	text = {RE},
	first = {reverse engineering (RE)},
}

\newglossaryentry{m2m}{
	name = {M2M},
	description = {machine to machine},
	text = {M2M},
	first = {machine to machine (M2M)}
}

\newglossaryentry{ip}{
	name = {IP},
	description = {intellectual property},
	text = {IP},
	first = {intellectual property (IP)}
}

\newglossaryentry{3pip}{
	name = {3PIP},
	description = {third-party intellectual property},
	text = {3PIP},
	first = {third-party intellectual property (3PIP)}
}

\newglossaryentry{dos}{
    name={DoS},
    description={denial of service},
    text = {DoS},
    first ={denial of service (DoS)}
}

\newglossaryentry{ddos}{
    name={DDoS},
    description={distributed denial of service},
    text = {DDoS},
    first ={distributed denial of service (DDoS)}
}

\newglossaryentry{soc}{
	name = {SoC},
	description = {system on chip},
	text = {SoC},
	first = {system on chip (SoC)},
	plural = {SoCs},
	firstplural = {systems on chip (SoCs)}
}

\newglossaryentry{sdr}{
    name={ SDR},
    description={software-defined radio},
    text={SDR},
    first={software-defined radio (SDR) }
}

\newglossaryentry{ic}{
    name={IC},
    description={integrated circuit},
    text={IC},
    first={integrated circuit (IC)},
	plural = {ICs},
	firstplural = {integrated circuits (ICs)}
}

\newglossaryentry{pcb}{
    name={PCB},
    description={printed circuit boards},
    text={PCB},
    first={printed circuit boards (PCB) }
}

\newglossaryentry{eda}{
    name={EDA},
    description={electronic design automation},
    text={EDA},
    first={electronic design automation (EDA) }
}

\newglossaryentry{iot}{
    name={IoT},
    description={internet of things},
    text={IoT},
    first={internet of things (IoT) }
}

\newglossaryentry{iiot}{
    name={IIoT},
    description={industrial internet of things},
    text={IIoT},
    first={industrial internet of things (IIoT) }
}

\newglossaryentry{sathv}{
	name={SATHV},
	description={software attacks targeting hardware vulnerabilities},
	text={SATHV},
	first={software attacks targeting hardware vulnerabilities (SATHV)}
}

\newglossaryentry{ml}{
	name={ML},
	description={machine learning},
	text={ML},
	first={machine learning (ML)}
}

\newglossaryentry{hls}{
	name={HLS},
	description={high-level synthesis},
	text={HLS},
	first={high-level synthesis (HLS)}
}

\newglossaryentry{feol}{
	name={FEOL},
	description={front end of line},
	text={FEOL},
	first={front end of line (FEOL)}
}

\newglossaryentry{beol}{
	name={BEOL},
	description={back end of line},
	text={BEOL},
	first={back end of line (BEOL)}
}

\newglossaryentry{obisa}{
	name={OBISA},
	description={obfuscated built-in self-authentication },
	text={OBISA},
	first={obfuscated built-in self-authentication (OBISA)}
}

\newglossaryentry{bisa}{
	name={BISA},
	description={ built-in self-authentication },
	text={BISA},
	first={built-in self-authentication (BISA)}
}

\newglossaryentry{hdl}{
	name={HDL},
	description={hardware description language },
	text={HDL},
	first={hardware description language (HDL)}
}

\newglossaryentry{ttt}{
	name={TTT},
	description={ticking timebomb Trojan },
	text={TTT},
	first={ticking timebomb Trojan (TTT)},
	plural = {TTTs}
}

\newglossaryentry{rtl}{
	name={RTL},
	description={register transfer level },
	text={RTL},
	first={register transfer level (RTL)}
}
\newglossaryentry{rtlift}{
	name={RTLIFT},
	description={register transfer level information flow tracking },
	text={RTLIFT},
	first={register transfer level information flow tracking (RTLIFT)}
}

\newglossaryentry{ift}{
	name={IFT},
	description={ information flow tracking },
	text={IFT},
	first={information flow tracking (IFT)}
}

\newglossaryentry{sca}{
	name={SCA},
	description={side channel analysis},
	text={SCA},
	first={side channel analysis (SCA)}
}

\newglossaryentry{sem}{
	name={SEM},
	description={scanning electron microscope},
	text={SEM},
	first={scanning electron microscope (SEM)}
}

\newglossaryentry{milp}{
	name={MILP},
	description={mixed-integer linear programming},
	text={MILP},
	first={mixed-integer linear programming (MILP)}
}

\newglossaryentry{esl}{
	name={ESL},
	description={electronic system level},
	text={ESL},
	first={electronic system level (ESL)}
}

\newglossaryentry{smt}{
	name={SMT},
	description={satisfiability modulo theory},
	text={SMT},
	first={satisfiability modulo theory(SMT)},
	plural = {SMTs},
	firstplural = {satisfiability modulo theories (SMTs)}
}

\newglossaryentry{aes}{
	name={AES},
	description={advanced encryption standard},
	text={AES},
	first={advanced encryption standard (AES)}
}

\newglossaryentry{qdi}{
	name={QDI},
	description={quasi-delay insensitive},
	text={QDI},
	first={quasi-delay insensitive (QDI)}
}
\newglossaryentry{tsb}{
	name={TSB},
	description={tri-state buffer},
	text={TSB},
	first={tri-state buffer (TSB)}
	plural = {TSBs},
	firstplural = {tri-state buffers (TSBs)}
}

\newglossaryentry{mcsoc}{
	name={MCSoC},
	description={multicore system on chip},
	text={MCSoC},
	first={multicore system on chip (MCSoC)}
	plural = {MCSoCs},
	firstplural = {multicore systems on chip (MCSoCs)}
}

\newglossaryentry{noc}{
	name={noc},
	description={network on chip},
	text={NoC},
	first={network on chip (NoC)}
}

\newglossaryentry{svm}{
	name={svm},
	description={support vector machine},
	text={SVM},
	first={support vector machine (SVM)}
}

\newglossaryentry{fpga}{
	name={fpga},
	description={field programmable gate array},
	text={FPGA},
	first={field programmable gate array (FPGA)}
}

\newglossaryentry{cad}{
	name={cad},
	description={computer-aided design},
	text={CAD},
	first={computer-aided design (CAD)}
}

\newglossaryentry{fsm}{
    name={fsm},
	description={finite state machine},
	text={FSM},
	first={finite state machine (FSM)}
}

\newglossaryentry{gnn}{
    name={gnn},
	description={graph neural network},
	text={GNN},
	first={graph neural network (GNN)}
}

\newglossaryentry{cfg}{
    name={cfg},
	description={control flow graph},
	text={CFG},
	first={control flow graph (CFG)}
}

\newglossaryentry{lut}{
    name={lut},
	description={look up table},
	text={LUT},
	first={look up table (LUT)}
}

\newglossaryentry{glift}{
    name={glift},
	description={gate-level information-flow tracking},
	text={GLIFT},
	first={gate-level information-flow tracking (GLIFT)}
}

\newglossaryentry{ics}{
    name={ics},
	description={industrial control systems},
	text={ICS},
	first={industrial control systems (ICS)}
}

\newglossaryentry{pchip}{
    name={pchip},
	description={proof-carrying hardware IP},
	text={PCHIP},
	first={proof-carrying hardware IP (PCHIP)}
}
\usepackage{booktabs}

\def\BibTeX{{\rm B\kern-.05em{\sc i\kern-.025em b}\kern-.08em
    T\kern-.1667em\lower.7ex\hbox{E}\kern-.125emX}}

\usepackage{todonotes,soul}

\soulregister\cite7
\soulregister\citep7
\soulregister\ref7
\soulregister\cref7
\soulregister\gls7
\soulregister\glspl7
\soulregister\Gls7
\soulregister\Glspl7
 
\begin{document}

\title{
Information Flow Tracking Methods for Protecting Cyber-Physical Systems against Hardware Trojans\\
 - a Survey -
}

\author{
\IEEEauthorblockN{ Sofia Maragkou}
\IEEEauthorblockA{\textit{Institute of Computer Technology, TU Wien} \\
\textit{Vienna University of Technology}\\
Gusshausstr. 27–29 / 384, 1040 Wien, Austria \\
sofia.maragkou@tuwien.ac.at}
\and
\IEEEauthorblockN{ Axel Jantsch}
\IEEEauthorblockA{\textit{Institute of Computer Technology, TU Wien} \\
\textit{Vienna University of Technology}\\
Gusshausstr. 27–29 / 384, 1040 Wien, Austria \\
axel.jantsch@tuwien.ac.at}
}

\maketitle

\begin{abstract}

%Industry 4.0
\Gls{cps} provide profitable surfaces for hardware attacks such as hardware Trojans.
Hardware Trojans can implement stealthy attacks such as leaking critical information, taking control of devices or harm humans.
In this article we review \gls{ift} methods for protecting \gls{cps} against hardware Trojans, and discuss their current limitations.
\gls{ift} methods are a promising approach for the detection of hardware Trojans in complex systems because the detection mechanism does not necessarily rely on potential Trojan behavior. However, in order to maximize the benefits research should focus more on black-box design models and consider real-world attack scenarios.

\end{abstract}


\glsresetall
\section{Introduction}
% CPPS and IIot 
% The complexity of those systems, increases the complexity of the security. 
%\Gls{cpps} \cite{Monostori2014} consist of autonomous and cooperative elements and sub-systems that are getting into connection with each other in application dependent ways, on and across all levels of production, from processes through machines up to production and logistics networks. 
%They are the basic elements of Industry 4.0.
%\Gls{cpps} that monitor, process, control and exchange sensitive information regarding the production.
%The alteration or the leakage of such information, or the alteration of the specified functionality of those systems can lead to great economic loss. 

% Hardware security
Hardware security began facing desultory challenges much later than software~\cite{Bhunia2019}. 
In 1996 a timing attack was published  \cite{Kocher1996} based on which sensitive information could be leaked from a cryptographic component. 
After this point, hardware security research became more systematic. 
From 2005 on \cite{XiaoxiaoWang2008, Bhunia2019} the field of hardware security has gained ground in the academic and the industrial world because it breaks the chain of trust known so far.

% Soc supply chain and threat models - short description
This chain of trust, from the hardware security perspective, begins at the \gls{ic} supply chain, where security vulnerabilities are formed by the needs of the market for fast and cheap technology.
The involvement of external entities in the design process and the internationally outsourced fabrication can create security breaches that can be even relevant for national security.
Design houses, in order to stay competitive, purchase \gls{3pip} cores from vendors and outsource the fabrication process without always verifying the returned product with respect to hardware security breaches.
The reason for that is that the verification of the purchased cores is an expensive process that requires resources and time.
Those \gls{ip} cores or chips are integrated and distributed to the customers.
Consequently, hardware security has to deal with attacks like \gls{ip} piracy, reverse engineering, counterfeit chips and hardware Trojans. 

In a real world scenario, when an \gls{ip} core is being purchased, the design house requests some design specification and the \gls{3pip} core vendor replies with the \gls{ip} core and the specifications of the \gls{ip} core. 
Throughout this information exchange, the only trusted part is the specification requested by the design house. 
The core in return, is considered untrusted and it is treated as black box. 
\Gls{ift} methods are a promising research direction for the detection of hardware Trojans because the verification can be based on the security specification of the application and not only on potentially malicious designs. 
Thus, the verification methods can be adapted based on the application.
In addition, those methods can be flexible regarding new attacks, and can be expandable in case of the alteration of the security specifications.

\subsection{Known Real World Attacks}

The real world hardware attacks are much more complicated than the attacks developed by the research community, since
real world attacks interact with different layers of the computing system and communicate with external systems over long distance. 
Compared to software, real world hardware attacks are less frequent.
The information that is publicly available about real world attacks is limited and specific details are rarely known to the public. 

% Kill switch
The real world attack that received most attention is the 2007 attack on a Syrian military radar~\cite{Adee2008,bbcisrael}. 
Even though the details were not officially revealed, all the indications suggest that the radar at a nuclear installation in Syria has been tampered. 
The attack took place in September 2007 and the nuclear installation was completely destroyed by Israeli bombing jets.
The Israeli jets, took off from southern Israel, crossed the Mediterranean Sea and the Syrian-Turkish borders and returned four hours later. 
The state of the art radars did not detect the jets, which raised suspicions for malicious alteration of their functionality.
Adee \cite{Adee2008} suspects a kill-switch or a backdoor in the off-the-shelf microprocessor that could block a bombing radar by an apparently remote command (trigger) without shutting down the whole system.
The difference of a kill switch and a backdoor is that the kill switch will shut off a specific chip when triggered, but a backdoor requires an intruder to implement the same effect.
The hypothesis of the kill switch is more likely and, in order to be implemented, requires the injection of extra logic. 
The HW and SW overhead for such an attack is very small which makes it hard to detect during testing, and the threat models discussed are the malicious designer and the malicious manufacturer. 
The microprocessor used remains unknown.  
This is not the only occasion where microprocessors including a kill switch have supposedly been used. 
According to anonymous sources from U.S defense department, it is known that a European chip maker is building microprocessors with a kill switch, and the French defense uses this technology for military applications.
% Supermicro
Undocumented microchips were found in the servers assembled by Supermicro \cite{Robertson2018, Mehta2020}, that implemented a doorway to the network of the original system, which incorporated memory, networking capacity and processing power. 
The attack aimed at leaking sensitive information over a long term.  

% Stuxnet
Stuxnet attack provides an example of the real world attack capabilities in the industrial environment \cite{Kushner2013}. 
Stuxnet is a worm that was introduced in the Microsoft Windows operating systems and it was targeting specific industrial control systems of Siemens which were used in Iran to run centrifuges. 
Until the target was found the worm was updating itself. 
The worm was compromising the targeted system by exploiting 'zero-day' vulnerabilities. 
After monitoring the operation, the worm was taking the control of the control system and it ran the centrifuges to the point of failure, returning false feedback to cover the failure until the damage was irreversible.

Hybrid attacks are very common in real world scenarios. The hybrid attacks can include hardware, software and firmware parts. Such an attack can be malicious software that exploits vulnerabilities of the hardware, damaging physical resources such as Stuxnet~\cite{Kushner2013}.

\subsection{Cyber-Physical Production Systems}\label{cpps}

% basic information
\Gls{cps} are sophisticated systems that combine physical and cyber units. 
They are used in many different applications and they are the fundamental units of the \gls{iot}.
Their functionality is based on the information exchange and the interaction with each other. 
According to the \cite{Yaacoub2020a}, the nature of the \gls{cps} makes them particularly sensitive to attacks, due to their heterogeneous nature, their reliance on data and their large scale. 

When those systems are integrated in the production environment then we refer to them as \gls{cpps}.
Often, \gls{cpps} expose a profitable surface to adversaries for hardware Trojan introduction, because they are complex, sophisticated structures that manage sensitive information with extend communication among them, which facilitates malicious functionality to stay hidden. 
Consequently, we consider securing the \gls{cpps} an emerging, critical issue.

According to \cite{Monostori2014}, the pyramid of the automation hierarchy known until recently, is decentralized in the concept of Industry 4.0. 
The information processing has been distributed in many control units which exchanging information with the goal to optimize the production process.
The control units have moved closer to the technical processes for efficiency, creating an interactive communication net among heterogeneous systems.
This creates the challenge to secure those components.

Assume that a hardware Trojan is included in one of the control units. In Industry 4.0 machines use \gls{m2m} communication for sensitive information exchange. 
That means that the authentication keys are stored and processed in the machines.
If the hardware Trojan leaks an authentication key to the adversary, she can take the control of the unit and possibly the control of the factory. 

In such a demanding environment the \gls{cpps} should stay consistent to the security requirements. 
Availability, integrity and confidentiality are only the basic guidelines of the properties that should be taken into consideration. 
The proof that the units of those systems comply to those properties and to more detailed ones can be achieved with \gls{ift} methods as we discuss in the next sections.

\subsection{Scope}

The scope of this report is to survey how \gls{ift} methodologies can secure \gls{cps} against hardware Trojan attacks and how those methods need to be further developed in order to be applicable in real world scenarios.

The remainder of this survey is organised as follows: Section \ref{ht} provides basic information about hardware Trojans. Section \ref{ift} refers to basic information for \gls{ift} methods and presents state of the art methodologies against information leakage. Finally, in section \ref{disc} we compare the \gls{ift} methods and we discuss future steps for research.

\section{Hardware Trojans}\label{ht}

Hardware Trojans are circuits with hidden, unspecified, malicious functionality that can be included in any phase of the \gls{ic} supply chain. 
In the environment of Industry 4.0, stealthy attacks like hardware Trojans can implement any kind of effect, including information leakage.
In this report we are interested in this kind of malicious activity.

\begin{figure}[!ht]
\centerline{\includegraphics[scale=0.4]{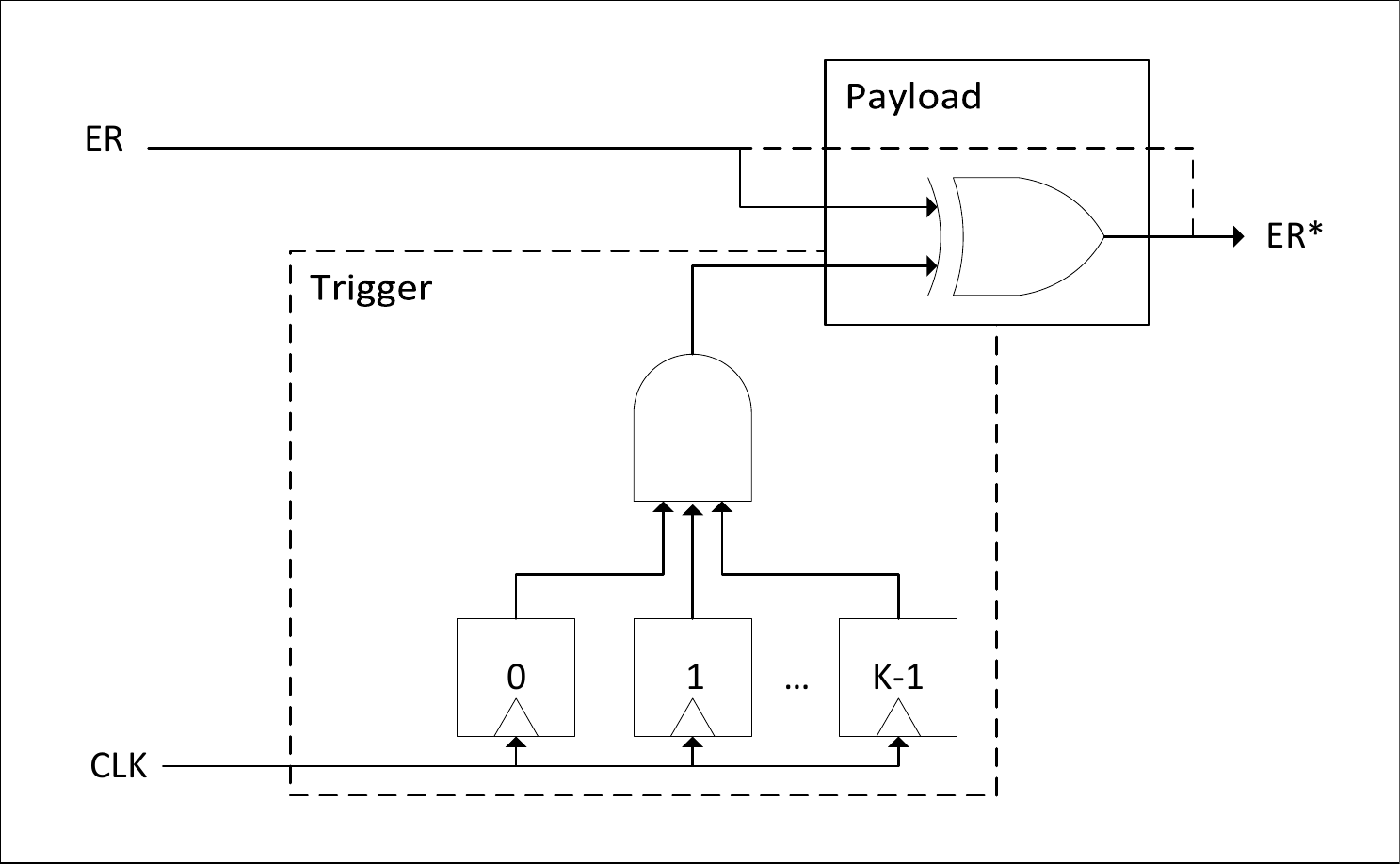}}
\caption{Time bomb hardware Trojan based on \cite{Chakraborty2009}}
\label{htf}
\end{figure}

Figure \ref{htf} shows a time bomb hardware Trojan from~\cite{Chakraborty2009}. 
This hardware Trojan is activated when the counter reaches the value $2^k-1$. 
When the trigger is activated, the output value at ER* becomes different from the initial signal ER.
The circuitry with the counter is the trigger and the circuitry that changes the value of the signal ER is the payload. 
This is a simplified example. More sophisticated mechanisms have been proposed from the research community like the Trojans mentioned above.

According to the taxonomy of R. Karri, J. Rajendran, K. Rosenfeld, M. Tehranipoor  \cite{Karri2010}, a hardware Trojan can be described by the insertion phase, the abstraction level, the activation mechanism (trigger), the effects (payload) and the location in the design.

\subsubsection{Insertion phase}
The earlier a hardware Trojan is introduced in the design the broader the range of its impact is and the lower the cost of the attack is.
For instance, assume that a third party vendor infects an \gls{ip} core with a hardware Trojan. 
This \gls{ip} core can be integrated in more than one design, increasing the number of infected systems.
On the other hand, the scenario of the malicious manufacturer is design-specific.
The attacker, in order to introduce a Trojan, should be aware of the design details which can be acquired by reverse engineering, a technique that needs special knowledge and is expensive in time and resources. 
Consequently, the phase of the hardware Trojan introduction, in combination with the value of the protected assets should be taken into consideration, during the development of countermeasures.

% Abstraction level
\subsubsection{Abstraction level}
Depending on the abstraction level of the design, a hardware Trojan can be injected at system level, at the development environment, at register-transfer level as soft \gls{ip} core, at gate level as firm \gls{ip} core, at transistor level as hard \gls{ip} core or at  the physical level.

% \Triggers
\subsubsection{Triggers}
There are hardware Trojans exploiting \textit{don't care conditions} for their trigger mechanisms~\cite{Hu2017}, or data patterns in specific memory addresses~\cite{ImtiazKhan2019}, or even dedicated input images~\cite{Zhao2019}. 
Some attacks have even more sophisticated triggers which are activated during the design flow, leaving no trigger signal to the possible detection algorithm \cite{Krieg2016, Ahmed2021}. 
% Since hardware designs are subjected to multiple tests before the distribution, the trigger has an important role, because it allows the payload part to stay hidden during testing, making the attack stealthy.

% Payload
\subsubsection{Payload}
The most common attacks realized by hardware Trojans are sensitive information leakage and \gls{dos} attacks. 
Other attacks can be functional alteration, downgrade performance, data corruption, circuit aging, chip destruction, etc. 

% Common targets
\subsubsection{Attack targets}
The most common targets for hardware Trojan attacks are memory elements \cite{ Hopkins2016,Kim2017,Hu2021,Nagarajan2019} and cryptographic components \cite{Liu2017,Schellenberg2018,Hu2017}.
However, there are many proposals for attacking cores such as UARTs \cite{Fern2015} or AXI4-bus interconnects \cite{Fern2016}, FPGA LUTs \cite{Krieg2016}, CPUs \cite{Gross2019,De2020,Zhang2020}, etc. 

% Resources required
\subsubsection{Resources required}
For the majority of the Trojans we study, the attacker needs knowledge of the design and access to it (e.g. bitstream \cite{Swierczynski2015}, netlists \cite{Cruz2018} or access to the design tools \cite{Pilato2019a, Krieg2016, Ahmed2021}).

% -----------------------------------------------------------------------

\section{Information flow tracking}\label{ift}
% Why IFT and why formal methods
The basic idea behind \gls{ift} methods is that they track the influence of information of a system during computation. 
In order to achieve that, they assign tags (usually binary values) for each of the data element of the design and they update the value of the tag based on the applied method and the applied security properties. 
The verification is achieved by the observation of the value of the tags. 

\Gls{ift} methods can be used with different verification techniques as it is described in the taxonomy in \cite{Hu2021b}. 
More specifically they can verify security properties through static methods like simulation, formal verification, emulation, and virtual prototyping or through dynamic methods like runtime monitoring techniques.

There are many \gls{ift} methods used with different verification techniques and at different abstraction levels and tackling different problems, since not all those methods address hardware Trojans.

Here in this paper we chose to present different \gls{ift} approaches and discuss their limitations and requirements. 
We present \gls{ift} static methods that tackle information leakage. 
Information leakage is the most common hardware Trojan effect and in the case of \gls{cps} it can cause economic loss or even set a human life in danger.

As we discussed earlier, the runtime monitoring methods can be expensive in resources, and the recovery from those attacks can be costly too. 
Based on that, we chose to focus on the static \gls{ift} verification methods.
Static \gls{ift} methods are applied in design-time, identifying the malicious behavior soon enough to minimize the recovery cost.
Moreover, they do not add overhead in the original designs resources.

% Provide examples of few methods
\section{IFT methods against hardware Trojans}

Many methodologies are using theorem proving to verify the information flow in the designs \cite{Bidmeshki2015, Guo2017a, Qin2019,Zhang2020a}.
In those methods the security properties are expressed as theorems and theorem proving tools such as Coq are used to verify them. 
In the \gls{pchip} framework~\cite{Bidmeshki2015} the \gls{ip} vendors are required to deliver the HDL code of the design with formal proofs that the code is according to some security properties predefined among the two parties.
For instance, such a property could describe that an instruction is allowed to access memory locations, which are defined in its op-code.
With the provided security tags to the signals \gls{pchip} can track the information flow in the design.
The disadvantage of theorem proving methods is the manual conversion of the HDL core to the theorem proving language and proof checking environment (e.g. Coq and CoqIDE).
Even though a conversion from HDL to Coq has been proposed \cite{Bidmeshki2015, Guo2017a}, theorem proving is far from an automated technique.

% Nahiyan2017 - black box approach but it is based on trigger condition Once we detect the presence of a Trojan
The approach proposed in \cite{Nahiyan2017} addresses black box models. 
It is based on information flow security (IFS) verification which detects violations of security properties. 
An asset is modeled as stuck-at-0 and stuck-at-1 faults and, by leveraging the automatic test pattern generation (ATPG), faults are searched for. When a fault is detected, it means that there is an information flow from the asset to observation points. Finally the trigger mechanisms is extracted. This methodology is based on the fact that the trigger mechanism is injected in the original circuit. 

% Ardeshiricham2017a --formal verification by standard tools
The tool Register Transfer Level Information Flow Tracking (RTLIFT)\cite{Ardeshiricham2017a}, can be applied directly to HDL code. 
Security tags (or labels) are assigned to every signal.
\Gls{rtlift} uses \gls{ift} logic to securely propagate the tags throughout the design. 
The functionality of the additional \gls{ift} logic depends on the precision required.
For instance, the output of an operation can be tainted when any of the inputs is tainted. If an untainted input influences the output to be untainted even though the other input is tainted, a false positive may occur. 
To avoid inaccuracies, the modules implementing the flow tracking logic take such cases into consideration.
Based on the required trade off between complexity and precision, different precision levels can be achieved.
Given the Verilog code, the control and the data flow precision flags (which define the required precision level), the tool generates a functionally equivalent Verilog code including \gls{ift} logic (IFT-Verilog code). 
The IFT-Verilog code is tested against the security properties requested for the design through simulation or formal verification. 
If the design passes this process, the extra logic is removed and the design is sent for fabrication. 
If it fails, the design has to be altered and to go through this process again. 

% Hu2016a - GLIFT is difficult due to the complexity  and produces false positives -- simulation
The methodology described in \cite{Hu2016a}, \gls{glift}, can detect hardware Trojans injected by malicious third-party vendors, that alter the functionality of the original circuit or leak sensitive information. 
According to \gls{glift}, each data bit is assigned to a security label.
This is implemented with additional tracking logic. 
It is up to the designers to define the security properties and use the \gls{glift} to verify the cores. 
For example, assume that the goal is to track the flow of a cryptographic key in order to ensure that it does not leak. 
The security labels of the keys will take the value 'confidential' and the security property that verifies that there is no leakage should ensure that no bit with 'confidential' label ends up in an output or memory with the label 'untrusted'.
Thus, this technique can identify violations of confidentiality and integrity and, hence, expose a hardware Trojan. 

 Both methods discussed above \cite{Hu2016a, Ardeshiricham2017a} face the problem of false positives results, which have to be resolved manually.

% Wang2018 - control and data flow graph with ift- based on trigger behavior -- feature matching
The method proposed by Wang et al. \cite{Wang2018}, called HLIFT, detects hardware Trojans based on the trigger behavior at \gls{rtl} with the use of control and data flow graphs (CDFG). 
The method can identify hardware Trojans that leak information through specific outputs pins or side channel, without functional modification and through unspecified output pins. 
This approach is based on a feature matching methodology that captures specific Trojan features. 
The features are based on three kind of Trojan triggers: always-on, immediate-on, sequential-on. 
This methodology can be divided in the predefinition flow and the application flow. 
During the predefinition flow, statement CDFGs are build based on already known infected \gls{rtl} designs.
Statement CDFGs are abstract, high-level and compact \gls{rtl} netlists.
That way unnecessary information is removed which decreases the complexity. 
\Gls{ift} is applied on the CDFGs and a list of Trojan \gls{ift} features is created.
At the application flow, the statement-level CDFG is extracted from the unknown \gls{rtl} design, and it is compared for matches with the list of the extracted Trojan features.  

The methodology proposed in \cite{Hassan2017} uses virtual prototyping (SystemC TLM 2.0) to identify information leakage or untrusted access. At the behavioral level there is a lack of design details. Thus, the security properties applied are very strict.
This can lead to false positives. This approach identifies the vulnerable paths and reports them to the user for inspection. Consequently, the inspection process is done manually, adding time overhead.    

The approach in \cite{Hu2018}, creates \gls{ift} models and optimizes them according to specific security properties. The security properties are compiled to security constraints and assertions, which are combined with the trimmed \gls{ift} model. Finally, the combination of the \gls{ift} model with the security constraints and assertions is verified through simulation, emulation or formal verification.   

In contrast to the methods presented above, the method in \cite{Liu2022} does not use any of the mentioned verification methods. 
The HDL code is converted to an abstract syntax tree (AST) to identify, track and localize anomaly behavior. The AST is converted to directed data-flow graph (DFG). This process automatically recognizes interaction between IP cores. By identifying the sink and the source signals, the tool detects vulnerabilities and finally locates the threats.

%%%%%%%%%%%%%%%%%%%%%%%%%%%%%%%%%%%

\begin{table}[htbp]

\caption{Static IFT methods - WB=white box, BB=black box, TP=theorem proving, MC=model checking, GL= gate level, SL=sequencial logic }
\begin{center}
\begin{tabular}{ccccl}

\toprule
Method                          & Abstraction  &  BB/  & Verification & Limitations  \\
                                & level        &    WB   &  method    &         \\
\midrule
\cite{Bidmeshki2015}            & RTL       & WB & TP      & based on \\
                                &      &  &       & conservative \\
                                &      &  &       &  rules \cite{Hu2021a}\\%\midrule
\cite{Qin2019}                  & GL & WB  & TP    & manual proof \\
                                & or RTL & &    & construction \\%\midrule
\cite{Zhang2020a}               & GL & WB    &  TP   & proof of genuine\\ 
                                &  &     &   &benchmark ,\\ 
                                &  &     &   &does not \\ 
                                &  &     &   & support SL \\ %\midrule
\cite{Guo2017a}                 & GL & WB & TP and MC    &high complexity,\\
                                &   &      &            &  false positives \\ %\midrule
\cite{Nahiyan2017}              & GL & BB    & partial scan ATPG    & based on \\
                                &  &    & analysis    & trigger condition\\%\midrule
\cite{Ardeshiricham2017a}       & RTL       & WB     &  simulation or    & challenged in \\
                                &        &    &  SAT solving    & complex \\ 
                                &        &           &          &  structures  \\%\midrule
\cite{Hu2016a}                  &   GL     &  WB   & simulation   & creates\\                                   &        &     &    & false positives\\  %\midrule
\cite{Wang2018}                 & RTL & WB & feature matching    & based on HT \\
                                &  &  & & features \\ %\midrule
\cite{Hassan2017}               &  behavioral & WB & virtual prototypes    & lack of design\\
                                &   &  & & details, \\
                                 &   &  & & manual inspection \\ %\midrule
\cite{Hu2018}                   &   RTL    &  WB  &  assertion based    & false positives\\
                                &    or GL   &    &   simulation   &\\
                                &       &    &  emulation    &\\ %\midrule
\cite{Liu2022}                  &    RTL &  WB  &   If-tracker   & false positives\\
\midrule
\end{tabular}
\label{tab1}
\end{center}
\end{table}

\section{Discussion and Conclusions}\label{disc}

The development of hardware Trojans is flourishing as they attract interest from the academia and industry.
%\glspl{cpps} provide a profitable attack surface. 
As countermeasures, \gls{ift} methodologies are very promising, because they can be flexible, adaptable and expandable based on the application. 

However, the \gls{ift} verification methodologies proposed so far, cannot be applied in real world scenarios. 
To the best of our knowledge, usually the purchased \gls{ip} cores are not in a white box form (usually the cores are purchased locked in order to avoid \gls{ip} piracy), or the specifications of the cores provided are considered untrusted.
Thus, the IP cores purchased are treated as black boxes.
That means that the internals of the purchased modules are unknown and can be leveraged from other layers of the systems (firmware or software) for potential attacks.

Thus, there is a need to explore more  \gls{ift} methods for black box designs without the usage of known hardware Trojan behaviors. 
The reason we suggest, that the known Trojan behaviors should not be taken into consideration is because the attackers want their Trojans to stay hidden, pushing the limits of the current known Trojan behaviors, in order to make them more stealthy.
A case in point is the development of trigger mechanisms. In recent years  there is the tendency to include the trigger mechanisms in the design flow, so that the detection methods  searching for trigger behaviors cannot detect them. 

On the other hand, methods that are based on security properties to identify unwanted or unspecified behavior in the designs seem more flexible with respect to unknown attacks.
However, the completeness of the security properties is an open problem.
Another issue is the definition of the security properties by the engineers.
Manual processes can result in vulnerabilities of the systems which can be leveraged by adversaries.

Identifying a hardware Trojan in a real world example can be very challenging, especially since the trigger mechanism is not necessarily part of the original design. 
In some concepts a fault, a vulnerability, or a backdoor may be no different from a well covered Trojan.
From the real world attacks we can conclude that the attack scenarios implemented are much more complete than the ones provided by academia. 
In the real world examples mentioned above we identify mechanisms that can communicate at great distance and can affect state of the art systems. 
The attacks were sophisticated enough with complicated mechanisms with more than negligible overhead.

It will be useful for the research community to explore more complicated attacks, across the levels of a computing system in order to facilitate corresponding countermeasures. 

% \section*{Acknowledgment}
% This work was enabled by TÜV AUSTRIA safeseclab Research Lab for Safety and Security in Industry,
% a research cooperation between TU Wien and TÜV AUSTRIA.

%%%%%%%%%%%%%%%%%%%%%%%%%%%%%%%%%%%%%%%%%%%%%%%%%%%%%%%%%%%%%%%%%%%%%%%%%%%%%%%%%%%%%%%%
% References
\bibliographystyle{IEEEtran}
\bibliography{literature.bib}
%%%%%%%%%%%%%%%%%%%%%%%%%%%%%%%%%%%%%%%%%%%%%%%%%%%%%%%%%%%%%%%%%%%%%%%%%%%%%%%%%%%%%%%%

\end{document}